\documentclass[fleqn,10pt]{wlscirep}
\usepackage[utf8]{inputenc}
\usepackage[T1]{fontenc}
\usepackage{float}

\usepackage{float}
\title{Resonance of Vector Vortex Beams in a Triangular Optical Cavity}

\author[1]{L. M. Rodrigues }
\author[1,+]{L. Marques Fagundes}
\author[1,2,+]{D. C. Salles}
\author[2,+]{G. H. dos Santos}
\author[2,+]{J. M. Kondo}
\author[2,+]{P. H. Souto Ribeiro}
\author[3,+]{A. Z. Khoury}
\author[2,+]{R. Medeiros de Ara\'ujo}
\affil[1]{Departamento de F\'isica, Universidade Federal de Santa Catarina, Florian\'opolis, SC, 88040-900, Brazil}
\affil[2]{Universit\'e Paris-Saclay,  Gif-sur-Yvette, 91190, France}
\affil[3]{Instituto de Física, Universidade Federal Fluminense, Niteroi, RJ, 24210-346, Brazil}

\affil[*]{lucasmarques\_fagundes@hotmail.com}

\affil[+]{these authors contributed equally to this work}

\begin{abstract}
We experimentally demonstrate resonance of first-order vector vortex beams (VVB) with a triangular optical cavity. We also show that, due to their symmetry properties, the VVBs commonly known as radial and azimuthal beams do not resonate at the same cavity length, which could be explored to use the triangular resonator as a mode sorter. In addition, an intracavity Pancharatnam phase shifter (PPS) is implemented in order to compensate for any birefringent phase that the cavity mirrors may introduce.
\end{abstract}
\begin{document}

\flushbottom
\maketitle
\thispagestyle{empty}

\section*{Introduction}
\label{sec:symmetry}
 Light beams with orbital angular momentum are also called vortex beams. Paradigmatic vortex beams
are the Laguerre Gaussian modes \cite{Allen92}. However, other kinds of vortex beams came up along with
the development of sophisticated optical approaches for generating, manipulating, and analyzing 
light fields \cite{gutierrez-vega2000, bandres2004, volke-sepulveda2002, abramochkin2010, ring2012, abramochkin2017, shen2018}. In fact, it was found that it was possible to construct stable optical modes
for which the polarization varies across the plane transverse to the propagation direction \cite{Rosales-Guzmán_2018, ndagano2018, Rosales-Guzmán_generation,zeilinger}. To this kind of light
beams was given the name Vector Vortex Beams (VVB) \cite{chen2003}. More than being a very curious type of
structured light, it has been shown that they exhibit nonseparability for polarization and spatial modes
\cite{Souza2007, Borges2010, milione2011, Pereira2014, Balthazar2016, Passos2018} and there are several practical applications for them \cite{abouraddy2006, cheng2009, roxworthy2010, fatemi2011, neugebauer2014, parigi2015, yuan2017, yuan2022, cheng2023}.

In most studies and applications, it is necessary to identify, measure or sort out vector vortex modes
and several methods have been developed so far \cite{ndagano2018, wang2016, wang2017}. Among these tasks, the hardest one is the realization of a mode sorter that could separate different vector vortex modes without 
destroying or even without imposing strong losses to them. This is a relevant task, because
these beams have a considerable potential for the implementation of quantum communication
schemes using single photons and squeezed light \cite{chille2016, vallone2014, dambrosio2012a, dambrosio2016}. For vortex scalar beams, mode sorters have been developed based on conformal transformations \cite{Berkhout2010} and optical cavities \cite{wei2020, Santos2021}.

Here, we take a step forward in the development of a triangular optical cavity that may work as a VVB mode sorter. First, we experimentally demonstrate resonance of VVBs within such cavity, introducing what we call an intracavity Pancharatnan Phase Shifter (PPS). Then, we demonstrate that two different types of VVB, namely radially and azimuthally polarized beams, resonate for different cavity lengths due to their symmetry/antisymmetry properties. Consequently, in principle, they can be sorted using such a cavity, provided it is kept in resonance with one of them. The scheme, explored here in a proof-of-principle experiment, represents a promising tool that could be very helpful in building quantum communication networks based on vortex vector beams.

\section*{Optical Cavities and First-order Vector Vortex Beams}

An optical cavity is an interferometric device that fold a light beam over itself to achieve resonance (constructive interference). The resonance condition is that the phase acquired by the beam after a cavity roundtrip is a multiple of $2\pi$. This phase depends on a list of factors. The two main factors are: i) the ratio between the roundtrip optical length $L$ and the wavelength $\lambda$ of light and ii) the beam spatial properties.

Mathematically, the phase accumulated over a cavity roundtrip can be written as a sum of contributions:
\begin{equation}
    \Delta\phi_\text{cav} = kL + \Delta\phi_\text{Gouy} + \Delta\phi_\text{M},
    \label{eq:roundtrip-phase}
\end{equation}
where $k$ is the wavenumber; $\Delta\phi_\text{Gouy}$ is the Gouy phase \cite{Gouy1890}, which plays an essential role in paraxial beams; and $\Delta\phi_\text{M}$ is the phase the mirrors may introduce in the beam, all summed up.

Hermite-Gaussian beams ($HG_{mn}$) and Laguerre-Gaussian beams ($LG_{\ell p}$) constitute two different and interesting bases to describe the propagation of paraxial beams. The mode order ($N = m+n$ or $N = |\ell|+2p$) and the Rayleigh length $z_0$ determine the amount of Gouy phase a beam will accumulate over a certain propagation distance $z$.

The first-order Hermite-Gauss modes ($HG_{10}$ and $HG_{01}$) accumulate the same phase after a roundtrip in a linear cavity composed of two mirrors, since these modes have the same Gouy phase. Therefore, they are resonant for the same cavity lengths and any superposition of them is transmitted by the output port of such cavity \cite{wei2020}. However, this is no longer the case for a triangular cavity composed of three mirrors, as we have shown in a previous experimental work \cite{Santos2021}. This type of cavity discerns optical modes based on their spatial symmetry. The explanation follows bellow.

\begin{figure}[h!]
\centering
   \includegraphics[width=0.9\textwidth]{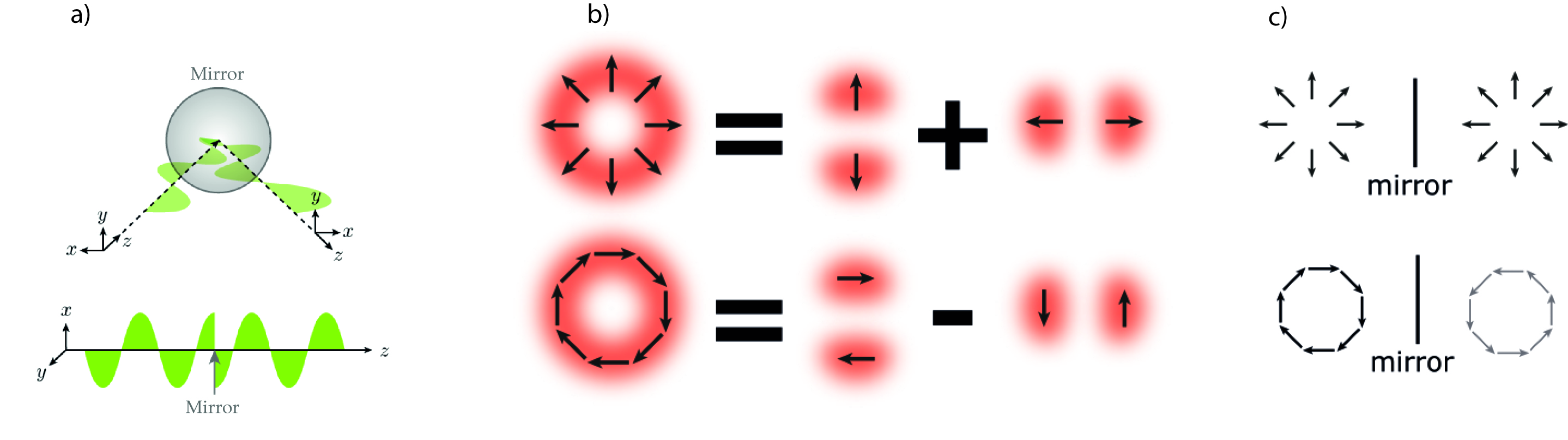}
   \caption{a) Inversion effect of the x-axis on the horizontal polarization (parallel to the plane of propagation) of a beam.b) VVB decomposition in the hermite-gaussian basis. First line: radial beam as in Eq. (\ref{eq:rad}); second line: azimuthal beam as in Eq. (\ref{eq:azim}). c) Symmetry and antisymmetry of radial and azimuthal VVBs upon reflexion.}
   \label{fig:symmetry}
\end{figure}

Take $z$ as the coordinate along the optical path inside the cavity and $x$ as the transverse coordinate whose axis is parallel to the incidence plane. A reflection causes the $z$ axis to invert its orientation (see Fig. \ref{fig:symmetry}a). So, if the beam's coordinate system is to keep its handedness, the $x$ axis must also change orientation. In this scenario, a reflection can be computed as a sign change in the horizontal coordinate and on the horizontal unit vector: $x \rightarrow -x$, $\hat{x} \rightarrow -\hat{x}$ \cite {Sasada2003}. The result is that a single reflection on a mirror leaves the $HG_{01}$ mode unchanged, while the $HG_{10}$ mode (the one with horizontally placed lobes) gets a minus sign, which is equivalent to a $\pi$ phase. This happens because the electric field of the $HG_{10}$ mode is an odd function on the variable $x$: the mode is antisymmetric. Therefore, in a triangular cavity, the three mirrors add no extra phase to $HG_{01}$ and a $\pi$ phase (=$3\pi \mod 2\pi$) to $HG_{10}$. When one solves the resonance condition $\Delta\phi_\text{cav}=2\pi$ using equation (\ref{eq:roundtrip-phase}) to find the resonance lengths, a $\pi$ shift in $\Delta\phi_\text{M}$ corresponds to a $\lambda/2$ shift in $L$. In conclusion, the resonance lengths for $HG_{01}$ and $HG_{10}$ should be separated by exactly half a wavelength.

When referring to paraxial beams, spatially homogeneous polarization is generally implicit. In this work, however, we are interested in the first-order Vector Vortex Beams, modes with inhomogeneous polarization profiles which are generated by linear superpositions of first-order Hermite-Gauss modes with orthogonal linear polarizations \cite{zhan2009}. Figure \ref{fig:symmetry}a illustrates this concept with the two cases we address throughout this paper: the radially polarized beam and the azimuthally polarized beam.

The discussion about reflection symmetry can now be extended to VVBs. Let us first analyze the matter by decomposing the VVBs as in Fig. \ref{fig:symmetry}b. The complex amplitude of the radial and azimuthal modes can be written, respectively, as 
\begin{align}
A^{\textit{rad}} &= HG_{01}\,\hat{y} + HG_{10}\,\hat{x}\,,\label{eq:rad}\\
A^{\textit{azim}} &= HG_{01}\,\hat{x} - HG_{10}\,\hat{y}\,\label{eq:azim}.
\end{align}
This shows that a single reflection will make $A^{\textit{rad}} \rightarrow A^{\textit{rad}}$ and $A^{\textit{azim}} \rightarrow -A^{\textit{azim}} = e^{i\pi}A^{\textit{azim}}$. The same conclusion may be drawn by analyzing the full polarization structure of the radial and azimuthal modes, which are symmetric and antisymmetric, respectively (see Fig. \ref{fig:symmetry}c). This $\pi$ phase difference leads to distinct resonance lengths for radial and azimuthal modes, exactly as discussed in the previous example ($HG_{01}$ versus $HG_{10}$).

Thus, a cavity with an odd number of mirrors is able to distinguish between radial and azimuthal beams, in the sense that they don't resonate simultaneously: whenever the radial beam is resonant (transmitted), the azimuthal beam is out of resonance (reflected). This is the case of a triangular cavity and that is precisely what we demonstrate experimentally below.

\section*{Geometric Phases}
Before diving into the experimental apparatus, it is worth presenting the concept of geometrical phases, which will be used to our advantage to ensure VVB resonance in the triangular optical cavity.

The notion of geometric phases was first introduced in the seminal work by Pancharatnam in 1956 \cite{Pancharatnam1956} in connection with cyclic transformations in the polarization state of a light beam. These cyclic transformations can be viewed as closed trajectories in the Poincaré sphere representation of the polarization state. The geometric phase acquired in the cycle is equal to half the solid angle $\Omega$ enclosed in the sphere and became known as Pancharatnam phase.

From a historical perspective, geometric phases have also been found in quantum systems. In 1984, Sir Michael Berry demonstrated their appearance in the adiabatic evolution of a quantum system state vector \cite{Berry1984} and this feature eventually became known as Berry phase. Later, this concept was extended to nonadiabatic evolutions \cite{Mukunda1993}. Geometric phases were shown to be a useful tool for quantum computation, where a conditional phase gate was demonstrated in both nuclear magnetic resonance \cite{Jones2000} and trapped ions \cite{Duan2001}. 

The Pancharatnam phase has also been demonstrated in the context of cyclic transformations on the transverse mode of a paraxial laser beam \cite{VanEnk1993,Galvez2003}. Besides its intrinsic beauty, the Pancharatnam phase has proved to be a useful tool for controlling the phase of a laser beam. Recently, it has been used to control the phase of an atom interferometer \cite{Decamps2017}. 

In our experiment, we apply the concept of geometric phase in the same context originally proposed by Pancharatnam, namely the polarization of light. We develop a tool that we call Pancharatnam Phase Shifter (PPS) and insert it inside the optical cavity in order to compensate for an unexpected polarization-dependent phase induced by the cavity mirrors. The reader will find in section ``Results and Discussion'' a detailed explanation on how the PPS works and what it is constituted of.

\section*{Experimental Setup}
\label{sec:exp}

The experiment described in this section was designed to test VVB resonance in a triangular optical cavity. In addition, observing the radial and azimuthal beams resonating in distinct cavity lengths would confirm the influence of the beams symmetry on the resonance length, predicted in previous section. 

\begin{figure}[H]
\centering
   \includegraphics[width=0.7\textwidth]{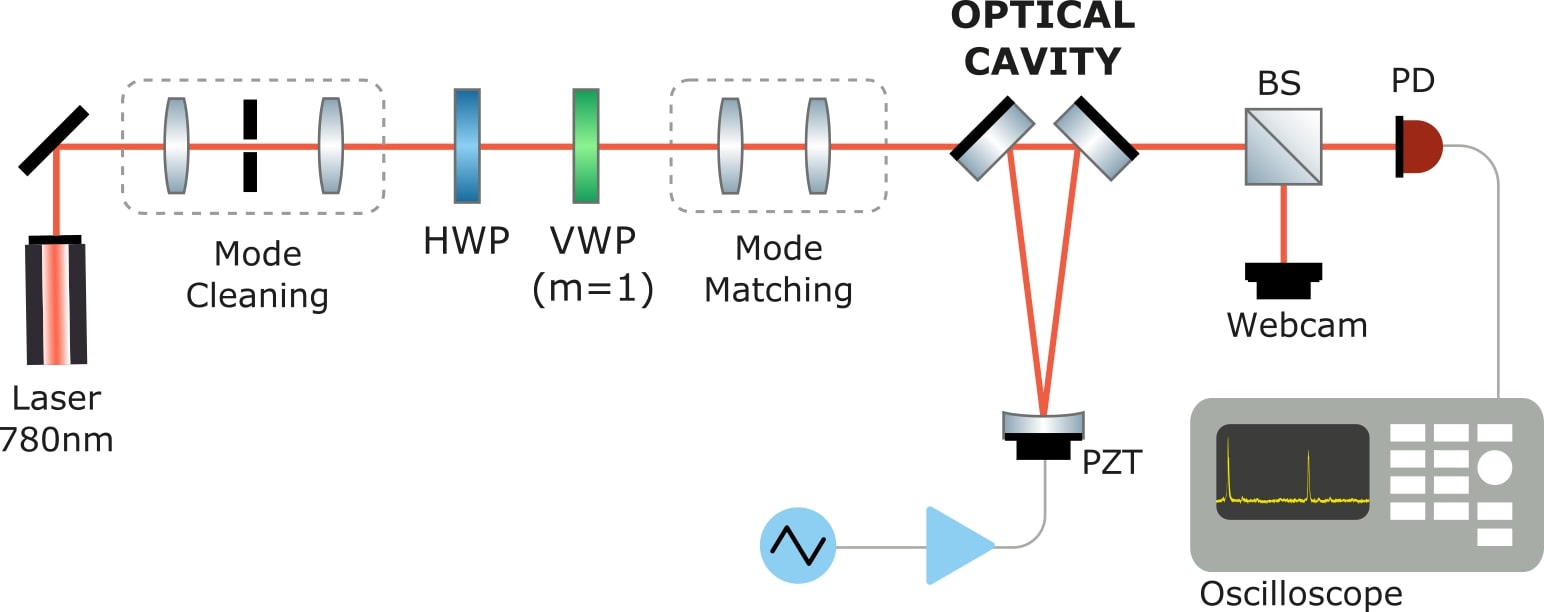}
   \caption{Experimental setup. HWP: half-wave plate; VWP: vortex waveplate; BS: beam-splitter; PD: photodiode; PZT: piezo actuator.}
   \label{fig:setup}
\end{figure}

Figure \ref{fig:setup} illustrates the experimental setup, which uses a homemade extended-cavity diode laser at 780 nm as the light source. A pinhole at the Fourier plane cleans the beam's spatial mode and a Spatial Light Modulator (not shown in the figure, for compactness) is used to finely adjust mode matching.

Once mode matching and beam/cavity alignment are optimized for a zero-order gaussian beam, a Vortex Wave Plate(VWP10-532 $m=1$, Thorlabs) is placed and centered on the optical path between the SLM and the mode matching lenses, with its fast axis aligned with the vertical. This VWP is a commercial wave plate that can be effectively understood as a ``mosaic'' of minuscule half-wave plates (HWP) with different orientations $\theta$ that equal half the azimuthal angle at the VWP plane. Also, a HWP is placed before the VWP to control the input polarization, determining which VVB is produced. If the input linear polarization is vertical (horizontal), a radial (azimuthal) beam is produced. A varying HWP angle will produce a superposition of radial and azimuthal with varying weights.

The generated VVB is then sent to the triangular cavity, composed of two identical partially reflective plane mirrors ($R = $ 96\% and 82\% for s- and p-polarization, respectively) and a concave mirror of high reflectance and a 200-mm radius of curvature. The cavity length (approximately 300 mm, roundtrip) is micrometrically scanned at $\sim$10 Hz using a piezoelectric transducer glued to the concave mirror and the resonance peaks are observed with an oscilloscope. Half of the cavity output intensity is sent to a simple webcam, in order to check the beam profile of the transmitted beams, peak by peak.

As will be discussed in the following section, resonance of VVB with our empty cavity was only nearly achieved, which prompted us to explore the insertion of a phase compensation system inside the cavity.

%%%%%%%%%%%%%%%%%%%%%%%%%%%%%%%%%%%%%%%%%%%%%%%
% FIG
%%%%%%%%%%%%%%%%%%%%%%%%%%%%%%%%%%%%%%%%%%%%%%%
%%%%%%%%%%%%%%%%%%%%%%%%%%%%%%%%%%%%%%%%%%%%%%%%

%%%%%%%%%%%%%%%%%%%%%%%%%%%%%%%%%%%%%%%%%%%%%%%
%  SECTION
%%%%%%%%%%%%%%%%%%%%%%%%%%%%%%%%%%%%%%%%%%%%%%%

\section*{Results and Discussion}\label{sec:results}

As discussed in Section Optical Cavities and First-order Vector Vortex Beams, the resonance of a first-order VVB beam is achieved if its Hermite-Gaussian components resonate simultaneously within the cavity. In the case of a radial (azimuthal) beam, these components are the modes $HG_{10}\,\hat{x}$ and $HG_{01}\,\hat{y}$ ($HG_{10}\,\hat{y}$ and $HG_{01}\,\hat{x}$).

\begin{figure}[H]
\centering
   \includegraphics[width=0.8\textwidth]{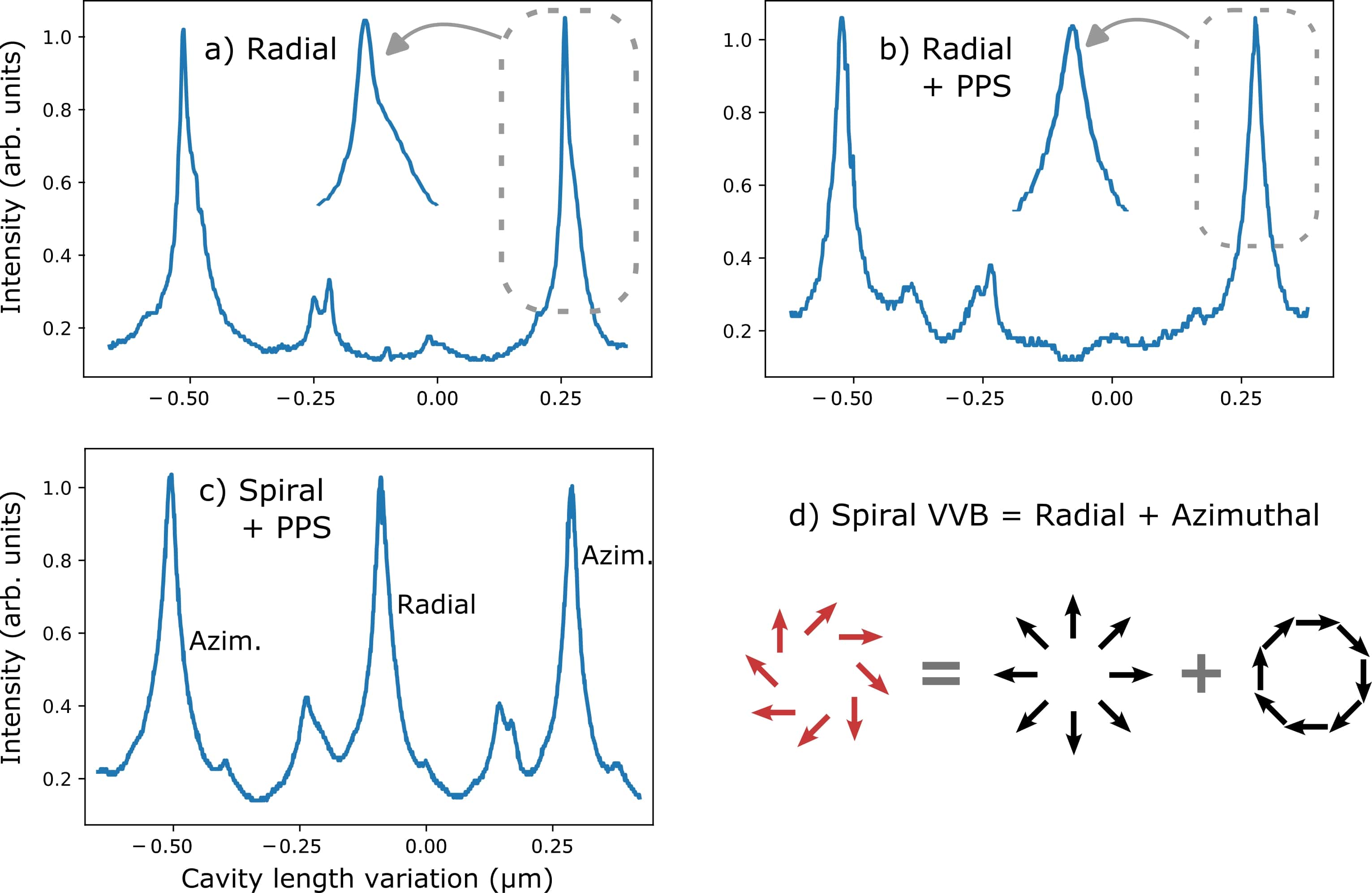} 
   \caption{Resonance peaks of a radially polarized beam for (a) an empty cavity and (b) for a cavity with Pancharatnam phase shifter (PPS); (c) Resonance peaks for a spirally polarized beam, which is decomposed (d) into a sum of a radial beam with an azimuthal beam. The origins of the horizontal axes are arbitrary and may drift from one graphic to another.}
   \label{fig:VVB-resonance}
\end{figure}

In Fig. \ref{fig:VVB-resonance}a, we can observe the intensity transmission of the cavity with respect to its length when a radial beam is sent as an input. The graphic shows two main resonant peaks separated by a wavelength $\lambda$ and small resonant peaks caused by spurious mode components populated at the VVB production with the vortex plate. As we zoom in at the peaks we notice a slight, although unambiguous, peak asymmetry, suggesting that polarization components H and V do not resonate simultaneously. Indeed, a clear signature of this resonance splitting is the fact that this peak may be decomposed into two displaced peaks with distinct widths (attesting for distinct finesses). We have fitted the asymmetric peak to a sum of two displaced lorentzian functions. The fitted displacement obtained with this technique was $(0.016 \pm 0.001)\times\lambda$, so that the measured splitting lies around $\approx\lambda/60$ in cavity length, or $\approx 2\pi/60$ in accumulated roundtrip phase.

%%%%%%%%%%%%%%%%%%%%%%%%%%%%%%%%%%%%%%%%%%%%%%%
% FIG
%%%%%%%%%%%%%%%%%%%%%%%%%%%%%%%%%%%%%%%%%%%%%%%
%%%%%%%%%%%%%%%%%%%%%%%%%%%%%%%%%%%%%%%%%%%%%%%% 

\begin{figure}[h!]
\centering
   \includegraphics[width=0.7\textwidth]{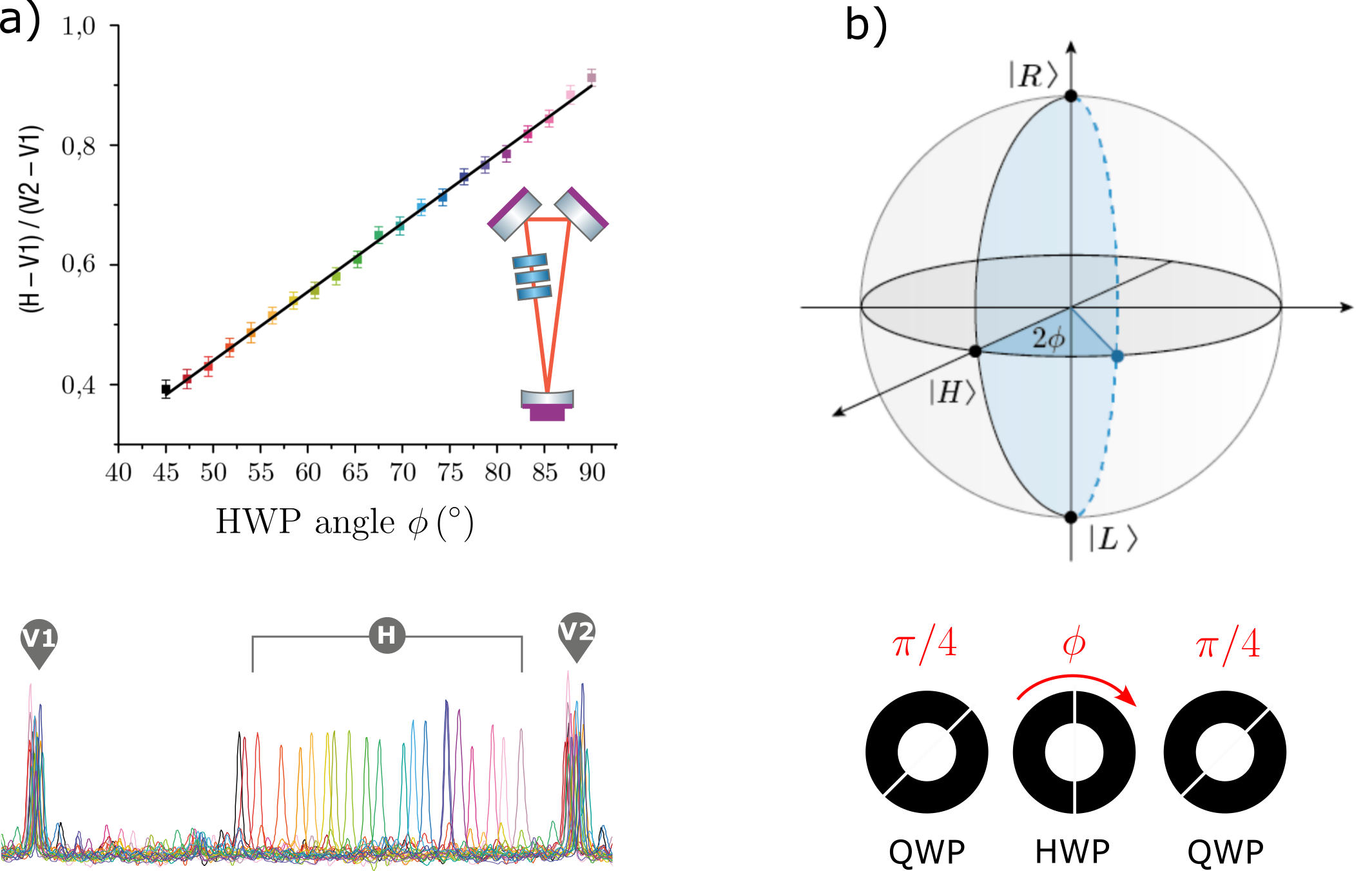} 
   \caption{(a) Effect of the intracavity phase adjustment on the resonance of horizontally and vertically polarized zero-order gaussian beams. In the vertical axis, H-V1 stands for the position of the H peak relative to the position of V1 peak (and similarly for V2). (b) Pancharatnam phase. First QWP at 45$^\circ$: H$\rightarrow$R. HWP at angle $\phi$: R$\rightarrow$L, dashed line. Second QWP at 45$^\circ$: L$\rightarrow$H.}
   \label{fig:pancharatnam2}
\end{figure}

This phase shift $\delta\varphi\approx 2\pi/60$ was unexpected. A simple explanation for it could be a modest birefringence on the thin films of the dielectric-coated cavity mirrors. In order to compensate for this phase shift and achieve simultaneous resonance of H and V components, we put a set of three waveplates (QWP + HWP + QWP) in the intracavity path, as shown in Fig. \ref{fig:pancharatnam2}a. This setup introduces an adjustable geometric phase to each polarization component without changing their direction, as explained below.

In our experiment, we compensate the relative phase acquired by the orthogonal polarization states in a cavity roundtrip, by introducing a Pancharatnam phase shifter (PPS) formed by a sequence of a quarter-waveplate fixed at $45^{\circ}$, a half-waveplate oriented at a variable angle $\phi\,$, and another quarter-waveplate fixed at $45^{\circ}$ placed inside the ring cavity. The crossed-polarized beams undergo cyclic polarization transformations as they pass through the phase shifter. Figure \ref{fig:pancharatnam2}b illustrates the transformation undergone by polarization H. It corresponds to a closed path in the Poincaré sphere with solid angle $\Omega=4\phi$. The acquired Pancharatnam phase should be therefore $\delta\varphi_H=\Omega/2=2\phi$. Polarization V acquires the same phase, but with opposite sign, by tracing a mirrored path at the other side of the sphere: $\delta\varphi_V=-\Omega/2=-2\phi$. 

In this manner, we were able to compensate for the cavity phase shift, by tuning the value of $\phi$ with the half-waveplate. The results of phase compensation for the radial beam are shown in Fig. \ref{fig:VVB-resonance}b, where the resonance peak is no longer asymmetric. This is the first indication that we have achieved resonance for the radially polarized VVB in a triangular cavity.

The graphic in Fig. \ref{fig:VVB-resonance}c shows that the radially and azimuthally polarized beams indeed resonate separately, i.e. at distinct cavity lengths, since they are symmetric and antisymmetric, respectively (see Fig. \ref{fig:symmetry}b). This graphic was obtained by sending a spirally polarized beam (Fig. \ref{fig:VVB-resonance}d) to the cavity, which is a superposition of radial and azimuthal modes with same weights. This superposition is produced by sending a diagonally polarized gaussian beam to the VWP.

%%%%%%%%%%%%%%%%%%%%%%%%%%%%%%%%%%%%%%%%%%%%%%%
% FIG
%%%%%%%%%%%%%%%%%%%%%%%%%%%%%%%%%%%%%%%%%%%%%%%
%%%%%%%%%%%%%%%%%%%%%%%%%%%%%%%%%%%%%%%%%%%%%%%%

\begin{figure}[H]
\centering
   \includegraphics[width=0.8\textwidth]{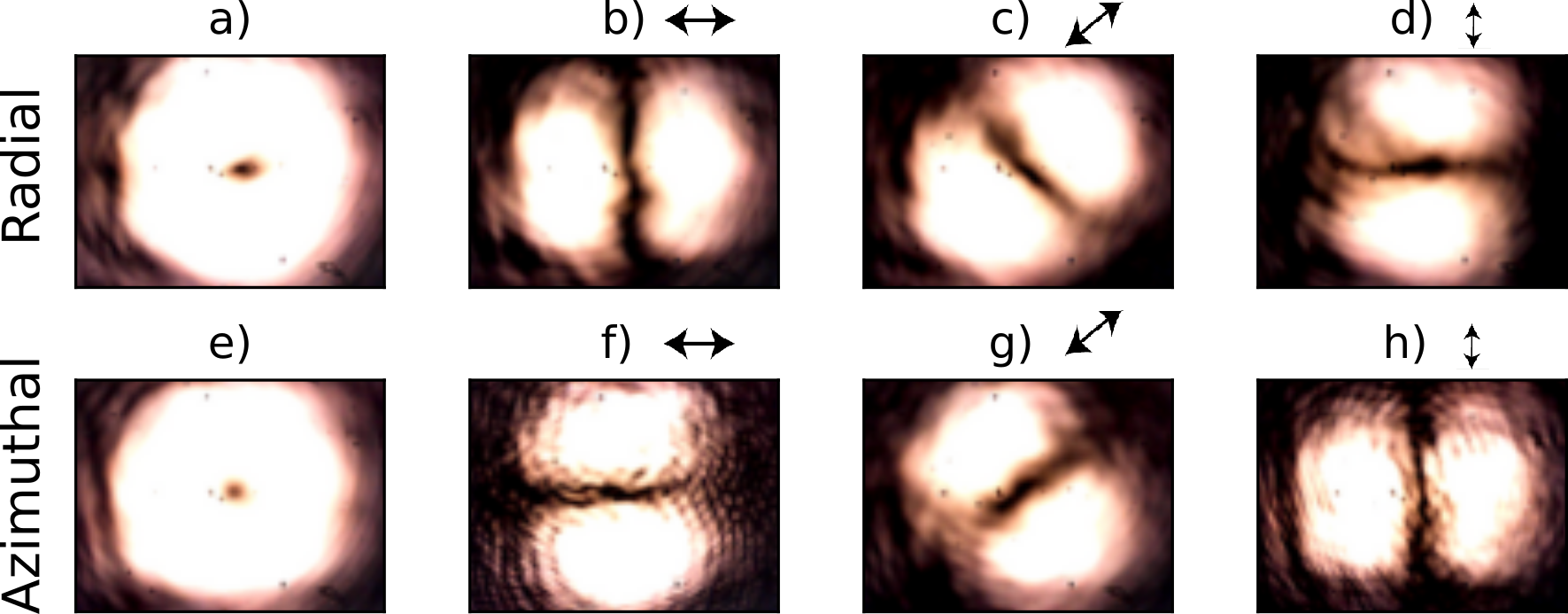}
   \caption{Intensity profiles of output resonant VVB beams. a) Transmitted radial beam and b)-d) its projections onto horizontal, diagonal ($+45^\circ$) and vertical polarizations; e) Transmitted azimuthal beam and f)-h) its projections.}
   \label{fig:transmission}
\end{figure}

In order to prove the resonant beams maintain their vectorial features after passing through the cavity, we captured their intensity profiles using a webcam and further analyzed their projections onto horizontal, diagonal, and vertical polarizations. This analysis was conducted by placing a polarizer at various orientations in front of the webcam. Although we do not have a locking system in place yet, the cavity is stable enough to manually keep it nearly in resonance for a few seconds while taking the pictures of the transmitted beam. The results are presented in Figure \ref{fig:transmission}, confirming our hypothesis. For instance, Fig. \ref{fig:transmission}c shows the projection of the transmitted radial beam onto the diagonal polarization, which corresponds to a first-order hermite-gaussian beam with lobes aligned to the diagonal, whereas the azimuthal beam, when projected onto the same direction, displays a mode oriented along the antidiagonal (Fig. \ref{fig:transmission}g).

We can note a slight deviation of the images shown in Figs. \ref{fig:transmission}c and \ref{fig:transmission}g from pure rotated Hermite-Gaussian modes, probably due to a small unbalance between the losses on the horizontal and vertical polarization components of the VVBs.

%%%%%%%%%%%%%%%%%%%%%%%%%%%%%%%%%%%%%%%%%%%%%%%
% FIG
%%%%%%%%%%%%%%%%%%%%%%%%%%%%%%%%%%%%%%%%%%%%%%%
%%%%%%%%%%%%%%%%%%%%%%%%%%%%%%%%%%%%%%%%%%%%%%%%

%%%%%%%%%%%%%%%%%%%%%%%%%%%%%%%%%%%%%%%%%%%%%%%
%  SECTION
%%%%%%%%%%%%%%%%%%%%%%%%%%%%%%%%%%%%%%%%%%%%%%%

\section*{Conclusion}
\label{sec:conc}

We have demonstrated that first-order vector vortex beams (VVB) resonate with a triangular optical cavity, even though they display non-uniform polarization. From a theoretical point of view, this fact becomes evident when one looks at the Hermite-Gauss decomposition of 1st-order VVBs. 

Experimentally, we have learned that an intracavity phase compensation may be necessary, depending on the dielectric mirror's features. In our experimental setup, we have observed a small unexpected displacement between horizontal and vertical resonance peaks, in addition to the expected displacement of half a free spectral range due the odd number of mirrors in the cavity. We hipothesize that this small displacement is due to the birefringence of dielectric thin layers that constitute the cavity mirrors (as in Ref. \cite{djevahirdjian2020}).

In order to compensate for this extra phase difference and ensure VVB resonance, we introduced a set of waveplates inside the cavity that adds an adjustable geometric phase, also refered to as Pancharatnam-Berry phase \cite{Pancharatnam1956}. We call this device a Pancharatnam phase shifter (PPS). Design improvements may be implemented in order to customize this device for future applications, making it more compact or mechanically robust.

Finally, we note that 1st-order VVB resonance should also occur for a linear cavity, although this has not been experimentally demonstrated in this paper. We highlight, though, that the triangular cavity has the unique feature of resonating with radial and azimuthal VVBs for distinct cavity lengths, suggesting this cavity could be used as a mode sorter for VVBs. The efficiency of this mode sorter is still to be assessed experimentally in future works, by electronically locking the cavity length on resonance and measuring the transmitted and reflected powers for each VVB input mode. Such a mode sorter, if/when available, could find applications, for example, in alignment-free Quantum Key Distribution protocols which uses vector vortex modes to construct mutually unbiased bases \cite{Aolita2007, Souza2008, dambrosio2012a}.

\bibliography{biblioVVBcavity}

\begin{thebibliography}{49}%
\makeatletter
\providecommand \@ifxundefined [1]{%
 \@ifx{#1\undefined}
}%
\providecommand \@ifnum [1]{%
 \ifnum #1\expandafter \@firstoftwo
 \else \expandafter \@secondoftwo
 \fi
}%
\providecommand \@ifx [1]{%
 \ifx #1\expandafter \@firstoftwo
 \else \expandafter \@secondoftwo
 \fi
}%
\providecommand \natexlab [1]{#1}%
\providecommand \enquote  [1]{``#1''}%
\providecommand \bibnamefont  [1]{#1}%
\providecommand \bibfnamefont [1]{#1}%
\providecommand \citenamefont [1]{#1}%
\providecommand \href@noop [0]{\@secondoftwo}%
\providecommand \href [0]{\begingroup \@sanitize@url \@href}%
\providecommand \@href[1]{\@@startlink{#1}\@@href}%
\providecommand \@@href[1]{\endgroup#1\@@endlink}%
\providecommand \@sanitize@url [0]{\catcode `\\12\catcode `\$12\catcode `\&12\catcode `\#12\catcode `\^12\catcode `\_12\catcode `\%12\relax}%
\providecommand \@@startlink[1]{}%
\providecommand \@@endlink[0]{}%
\providecommand \url  [0]{\begingroup\@sanitize@url \@url }%
\providecommand \@url [1]{\endgroup\@href {#1}{\urlprefix }}%
\providecommand \urlprefix  [0]{URL }%
\providecommand \Eprint [0]{\href }%
\providecommand \doibase [0]{https://doi.org/}%
\providecommand \selectlanguage [0]{\@gobble}%
\providecommand \bibinfo  [0]{\@secondoftwo}%
\providecommand \bibfield  [0]{\@secondoftwo}%
\providecommand \translation [1]{[#1]}%
\providecommand \BibitemOpen [0]{}%
\providecommand \bibitemStop [0]{}%
\providecommand \bibitemNoStop [0]{.\EOS\space}%
\providecommand \EOS [0]{\spacefactor3000\relax}%
\providecommand \BibitemShut  [1]{\csname bibitem#1\endcsname}%
\let\auto@bib@innerbib\@empty
%</preamble>
\bibitem [{\citenamefont {Allen}\ \emph {et~al.}(1992)\citenamefont {Allen}, \citenamefont {Beijersbergen}, \citenamefont {Spreeuw},\ and\ \citenamefont {Woerdman}}]{Allen92}%
  \BibitemOpen
  \bibfield  {author} {\bibinfo {author} {\bibfnamefont {L.}~\bibnamefont {Allen}}, \bibinfo {author} {\bibfnamefont {M.~W.}\ \bibnamefont {Beijersbergen}}, \bibinfo {author} {\bibfnamefont {R.~J.~C.}\ \bibnamefont {Spreeuw}},\ and\ \bibinfo {author} {\bibfnamefont {J.~P.}\ \bibnamefont {Woerdman}},\ }\bibfield  {title} {\bibinfo {title} {Orbital angular momentum of light and the transformation of laguerre-gaussian laser modes},\ }\href {https://doi.org/10.1103/PhysRevA.45.8185} {\bibfield  {journal} {\bibinfo  {journal} {Phys. Rev. A}\ }\textbf {\bibinfo {volume} {45}},\ \bibinfo {pages} {8185} (\bibinfo {year} {1992})}\BibitemShut {NoStop}%
\bibitem [{\citenamefont {{Guti{\'e}rrez-Vega}}\ \emph {et~al.}(2000)\citenamefont {{Guti{\'e}rrez-Vega}}, \citenamefont {{Iturbe-Castillo}},\ and\ \citenamefont {{Ch{\'a}vez-Cerda}}}]{gutierrez-vega2000}%
  \BibitemOpen
  \bibfield  {author} {\bibinfo {author} {\bibfnamefont {J.~C.}\ \bibnamefont {{Guti{\'e}rrez-Vega}}}, \bibinfo {author} {\bibfnamefont {M.~D.}\ \bibnamefont {{Iturbe-Castillo}}},\ and\ \bibinfo {author} {\bibfnamefont {S.}~\bibnamefont {{Ch{\'a}vez-Cerda}}},\ }\bibfield  {title} {\bibinfo {title} {Alternative formulation for invariant optical fields: {{Mathieu}} beams},\ }\href {https://doi.org/10.1364/OL.25.001493} {\bibfield  {journal} {\bibinfo  {journal} {Optics Letters}\ }\textbf {\bibinfo {volume} {25}},\ \bibinfo {pages} {1493} (\bibinfo {year} {2000})}\BibitemShut {NoStop}%
\bibitem [{\citenamefont {Bandres}\ and\ \citenamefont {{Guti{\'e}rrez-Vega}}(2004)}]{bandres2004}%
  \BibitemOpen
  \bibfield  {author} {\bibinfo {author} {\bibfnamefont {M.~A.}\ \bibnamefont {Bandres}}\ and\ \bibinfo {author} {\bibfnamefont {J.~C.}\ \bibnamefont {{Guti{\'e}rrez-Vega}}},\ }\bibfield  {title} {\bibinfo {title} {Ince\textendash{{Gaussian}} beams},\ }\href {https://doi.org/10.1364/OL.29.000144} {\bibfield  {journal} {\bibinfo  {journal} {Optics Letters}\ }\textbf {\bibinfo {volume} {29}},\ \bibinfo {pages} {144} (\bibinfo {year} {2004})}\BibitemShut {NoStop}%
\bibitem [{\citenamefont {{Volke-Sepulveda}}\ \emph {et~al.}(2002)\citenamefont {{Volke-Sepulveda}}, \citenamefont {{Garc{\'e}s-Ch{\'a}vez}}, \citenamefont {{Ch{\'a}vez-Cerda}}, \citenamefont {Arlt},\ and\ \citenamefont {Dholakia}}]{volke-sepulveda2002}%
  \BibitemOpen
  \bibfield  {author} {\bibinfo {author} {\bibfnamefont {K.}~\bibnamefont {{Volke-Sepulveda}}}, \bibinfo {author} {\bibfnamefont {V.}~\bibnamefont {{Garc{\'e}s-Ch{\'a}vez}}}, \bibinfo {author} {\bibfnamefont {S.}~\bibnamefont {{Ch{\'a}vez-Cerda}}}, \bibinfo {author} {\bibfnamefont {J.}~\bibnamefont {Arlt}},\ and\ \bibinfo {author} {\bibfnamefont {K.}~\bibnamefont {Dholakia}},\ }\bibfield  {title} {\bibinfo {title} {Orbital angular momentum of a high-order {{Bessel}} light beam},\ }\href {https://doi.org/10.1088/1464-4266/4/2/373} {\bibfield  {journal} {\bibinfo  {journal} {Journal of Optics B: Quantum and Semiclassical Optics}\ }\textbf {\bibinfo {volume} {4}},\ \bibinfo {pages} {S82} (\bibinfo {year} {2002})}\BibitemShut {NoStop}%
\bibitem [{\citenamefont {Abramochkin}\ and\ \citenamefont {Volostnikov}(2010)}]{abramochkin2010}%
  \BibitemOpen
  \bibfield  {author} {\bibinfo {author} {\bibfnamefont {E.~G.}\ \bibnamefont {Abramochkin}}\ and\ \bibinfo {author} {\bibfnamefont {V.~G.}\ \bibnamefont {Volostnikov}},\ }\bibfield  {title} {\bibinfo {title} {Generalized {{Hermite-Laguerre-Gauss}} beams},\ }\href {https://doi.org/10.3103/S1541308X10010036} {\bibfield  {journal} {\bibinfo  {journal} {Physics of Wave Phenomena}\ }\textbf {\bibinfo {volume} {18}},\ \bibinfo {pages} {14} (\bibinfo {year} {2010})}\BibitemShut {NoStop}%
\bibitem [{\citenamefont {Ring}\ \emph {et~al.}(2012)\citenamefont {Ring}, \citenamefont {Lindberg}, \citenamefont {Mourka}, \citenamefont {Mazilu}, \citenamefont {Dholakia},\ and\ \citenamefont {Dennis}}]{ring2012}%
  \BibitemOpen
  \bibfield  {author} {\bibinfo {author} {\bibfnamefont {J.~D.}\ \bibnamefont {Ring}}, \bibinfo {author} {\bibfnamefont {J.}~\bibnamefont {Lindberg}}, \bibinfo {author} {\bibfnamefont {A.}~\bibnamefont {Mourka}}, \bibinfo {author} {\bibfnamefont {M.}~\bibnamefont {Mazilu}}, \bibinfo {author} {\bibfnamefont {K.}~\bibnamefont {Dholakia}},\ and\ \bibinfo {author} {\bibfnamefont {M.~R.}\ \bibnamefont {Dennis}},\ }\bibfield  {title} {\bibinfo {title} {Auto-focusing and self-healing of {{Pearcey}} beams},\ }\href {https://doi.org/10.1364/OE.20.018955} {\bibfield  {journal} {\bibinfo  {journal} {Optics Express}\ }\textbf {\bibinfo {volume} {20}},\ \bibinfo {pages} {18955} (\bibinfo {year} {2012})}\BibitemShut {NoStop}%
\bibitem [{\citenamefont {Abramochkin}\ and\ \citenamefont {Alieva}(2017)}]{abramochkin2017}%
  \BibitemOpen
  \bibfield  {author} {\bibinfo {author} {\bibfnamefont {E.}~\bibnamefont {Abramochkin}}\ and\ \bibinfo {author} {\bibfnamefont {T.}~\bibnamefont {Alieva}},\ }\bibfield  {title} {\bibinfo {title} {Closed-form expression for mutual intensity evolution of {{Hermite}}\textendash{{Laguerre}}\textendash{{Gaussian Schell-model}} beams},\ }\href {https://doi.org/10.1364/OL.42.004032} {\bibfield  {journal} {\bibinfo  {journal} {Optics Letters}\ }\textbf {\bibinfo {volume} {42}},\ \bibinfo {pages} {4032} (\bibinfo {year} {2017})}\BibitemShut {NoStop}%
\bibitem [{\citenamefont {Shen}\ \emph {et~al.}(2018)\citenamefont {Shen}, \citenamefont {Wan}, \citenamefont {Meng}, \citenamefont {Fu},\ and\ \citenamefont {Gong}}]{shen2018}%
  \BibitemOpen
  \bibfield  {author} {\bibinfo {author} {\bibfnamefont {Y.}~\bibnamefont {Shen}}, \bibinfo {author} {\bibfnamefont {Z.}~\bibnamefont {Wan}}, \bibinfo {author} {\bibfnamefont {Y.}~\bibnamefont {Meng}}, \bibinfo {author} {\bibfnamefont {X.}~\bibnamefont {Fu}},\ and\ \bibinfo {author} {\bibfnamefont {M.}~\bibnamefont {Gong}},\ }\bibfield  {title} {\bibinfo {title} {Polygonal {{Vortex Beams}}},\ }\href {https://doi.org/10.1109/JPHOT.2018.2858845} {\bibfield  {journal} {\bibinfo  {journal} {IEEE Photonics Journal}\ }\textbf {\bibinfo {volume} {10}},\ \bibinfo {pages} {1} (\bibinfo {year} {2018})}\BibitemShut {NoStop}%
\bibitem [{\citenamefont {Chen}\ \emph {et~al.}(2003)\citenamefont {Chen}, \citenamefont {Huang}, \citenamefont {Lai},\ and\ \citenamefont {Lan}}]{chen2003}%
  \BibitemOpen
  \bibfield  {author} {\bibinfo {author} {\bibfnamefont {Y.~F.}\ \bibnamefont {Chen}}, \bibinfo {author} {\bibfnamefont {K.~F.}\ \bibnamefont {Huang}}, \bibinfo {author} {\bibfnamefont {H.~C.}\ \bibnamefont {Lai}},\ and\ \bibinfo {author} {\bibfnamefont {Y.~P.}\ \bibnamefont {Lan}},\ }\bibfield  {title} {\bibinfo {title} {Observation of {{Vector Vortex Lattices}} in {{Polarization States}} of an {{Isotropic Microcavity Laser}}},\ }\href {https://doi.org/10.1103/PhysRevLett.90.053904} {\bibfield  {journal} {\bibinfo  {journal} {Physical Review Letters}\ }\textbf {\bibinfo {volume} {90}},\ \bibinfo {pages} {053904} (\bibinfo {year} {2003})}\BibitemShut {NoStop}%
\bibitem [{\citenamefont {Souza}\ \emph {et~al.}(2007)\citenamefont {Souza}, \citenamefont {Huguenin}, \citenamefont {Milman},\ and\ \citenamefont {Khoury}}]{Souza2007}%
  \BibitemOpen
  \bibfield  {author} {\bibinfo {author} {\bibfnamefont {C.}~\bibnamefont {Souza}}, \bibinfo {author} {\bibfnamefont {J.}~\bibnamefont {Huguenin}}, \bibinfo {author} {\bibfnamefont {P.}~\bibnamefont {Milman}},\ and\ \bibinfo {author} {\bibfnamefont {A.}~\bibnamefont {Khoury}},\ }\bibfield  {title} {\bibinfo {title} {Topological phase for spin-orbit transformations on a laser beam},\ }\href@noop {} {\bibfield  {journal} {\bibinfo  {journal} {Physical review letters}\ }\textbf {\bibinfo {volume} {99}},\ \bibinfo {pages} {160401} (\bibinfo {year} {2007})}\BibitemShut {NoStop}%
\bibitem [{\citenamefont {Borges}\ \emph {et~al.}(2010)\citenamefont {Borges}, \citenamefont {Hor-Meyll}, \citenamefont {Huguenin},\ and\ \citenamefont {Khoury}}]{Borges2010}%
  \BibitemOpen
  \bibfield  {author} {\bibinfo {author} {\bibfnamefont {C.}~\bibnamefont {Borges}}, \bibinfo {author} {\bibfnamefont {M.}~\bibnamefont {Hor-Meyll}}, \bibinfo {author} {\bibfnamefont {J.}~\bibnamefont {Huguenin}},\ and\ \bibinfo {author} {\bibfnamefont {A.}~\bibnamefont {Khoury}},\ }\bibfield  {title} {\bibinfo {title} {Bell-like inequality for the spin-orbit separability of a laser beam},\ }\href@noop {} {\bibfield  {journal} {\bibinfo  {journal} {Physical Review A}\ }\textbf {\bibinfo {volume} {82}},\ \bibinfo {pages} {033833} (\bibinfo {year} {2010})}\BibitemShut {NoStop}%
\bibitem [{\citenamefont {Milione}\ \emph {et~al.}(2011)\citenamefont {Milione}, \citenamefont {Sztul}, \citenamefont {Nolan},\ and\ \citenamefont {Alfano}}]{milione2011}%
  \BibitemOpen
  \bibfield  {author} {\bibinfo {author} {\bibfnamefont {G.}~\bibnamefont {Milione}}, \bibinfo {author} {\bibfnamefont {H.~I.}\ \bibnamefont {Sztul}}, \bibinfo {author} {\bibfnamefont {D.~A.}\ \bibnamefont {Nolan}},\ and\ \bibinfo {author} {\bibfnamefont {R.~R.}\ \bibnamefont {Alfano}},\ }\bibfield  {title} {\bibinfo {title} {Higher-{{Order Poincar}}\textbackslash 'e {{Sphere}}, {{Stokes Parameters}}, and the {{Angular Momentum}} of {{Light}}},\ }\href {https://doi.org/10.1103/PhysRevLett.107.053601} {\bibfield  {journal} {\bibinfo  {journal} {Physical Review Letters}\ }\textbf {\bibinfo {volume} {107}},\ \bibinfo {pages} {053601} (\bibinfo {year} {2011})}\BibitemShut {NoStop}%
\bibitem [{\citenamefont {Pereira}\ \emph {et~al.}(2014)\citenamefont {Pereira}, \citenamefont {Khoury},\ and\ \citenamefont {Dechoum}}]{Pereira2014}%
  \BibitemOpen
  \bibfield  {author} {\bibinfo {author} {\bibfnamefont {L.}~\bibnamefont {Pereira}}, \bibinfo {author} {\bibfnamefont {A.}~\bibnamefont {Khoury}},\ and\ \bibinfo {author} {\bibfnamefont {K.}~\bibnamefont {Dechoum}},\ }\bibfield  {title} {\bibinfo {title} {Quantum and classical separability of spin-orbit laser modes},\ }\href@noop {} {\bibfield  {journal} {\bibinfo  {journal} {Physical Review A}\ }\textbf {\bibinfo {volume} {90}},\ \bibinfo {pages} {053842} (\bibinfo {year} {2014})}\BibitemShut {NoStop}%
\bibitem [{\citenamefont {Balthazar}\ \emph {et~al.}(2016)\citenamefont {Balthazar}, \citenamefont {Souza}, \citenamefont {Caetano}, \citenamefont {Galv{\~a}o}, \citenamefont {Huguenin},\ and\ \citenamefont {Khoury}}]{Balthazar2016}%
  \BibitemOpen
  \bibfield  {author} {\bibinfo {author} {\bibfnamefont {W.}~\bibnamefont {Balthazar}}, \bibinfo {author} {\bibfnamefont {C.}~\bibnamefont {Souza}}, \bibinfo {author} {\bibfnamefont {D.}~\bibnamefont {Caetano}}, \bibinfo {author} {\bibfnamefont {E.}~\bibnamefont {Galv{\~a}o}}, \bibinfo {author} {\bibfnamefont {J.}~\bibnamefont {Huguenin}},\ and\ \bibinfo {author} {\bibfnamefont {A.}~\bibnamefont {Khoury}},\ }\bibfield  {title} {\bibinfo {title} {Tripartite nonseparability in classical optics},\ }\href@noop {} {\bibfield  {journal} {\bibinfo  {journal} {Optics letters}\ }\textbf {\bibinfo {volume} {41}},\ \bibinfo {pages} {5797} (\bibinfo {year} {2016})}\BibitemShut {NoStop}%
\bibitem [{\citenamefont {Passos}\ \emph {et~al.}(2018)\citenamefont {Passos}, \citenamefont {Balthazar}, \citenamefont {de~Barros}, \citenamefont {Souza}, \citenamefont {Khoury},\ and\ \citenamefont {Huguenin}}]{Passos2018}%
  \BibitemOpen
  \bibfield  {author} {\bibinfo {author} {\bibfnamefont {M.}~\bibnamefont {Passos}}, \bibinfo {author} {\bibfnamefont {W.}~\bibnamefont {Balthazar}}, \bibinfo {author} {\bibfnamefont {J.~A.}\ \bibnamefont {de~Barros}}, \bibinfo {author} {\bibfnamefont {C.}~\bibnamefont {Souza}}, \bibinfo {author} {\bibfnamefont {A.}~\bibnamefont {Khoury}},\ and\ \bibinfo {author} {\bibfnamefont {J.}~\bibnamefont {Huguenin}},\ }\bibfield  {title} {\bibinfo {title} {Classical analog of quantum contextuality in spin-orbit laser modes},\ }\href@noop {} {\bibfield  {journal} {\bibinfo  {journal} {Physical Review A}\ }\textbf {\bibinfo {volume} {98}},\ \bibinfo {pages} {062116} (\bibinfo {year} {2018})}\BibitemShut {NoStop}%
\bibitem [{\citenamefont {Abouraddy}\ and\ \citenamefont {Toussaint}(2006)}]{abouraddy2006}%
  \BibitemOpen
  \bibfield  {author} {\bibinfo {author} {\bibfnamefont {A.~F.}\ \bibnamefont {Abouraddy}}\ and\ \bibinfo {author} {\bibfnamefont {K.~C.}\ \bibnamefont {Toussaint}},\ }\bibfield  {title} {\bibinfo {title} {Three-{{Dimensional Polarization Control}} in {{Microscopy}}},\ }\href {https://doi.org/10.1103/PhysRevLett.96.153901} {\bibfield  {journal} {\bibinfo  {journal} {Physical Review Letters}\ }\textbf {\bibinfo {volume} {96}},\ \bibinfo {pages} {153901} (\bibinfo {year} {2006})}\BibitemShut {NoStop}%
\bibitem [{\citenamefont {Cheng}\ \emph {et~al.}(2009)\citenamefont {Cheng}, \citenamefont {Haus},\ and\ \citenamefont {Zhan}}]{cheng2009}%
  \BibitemOpen
  \bibfield  {author} {\bibinfo {author} {\bibfnamefont {W.}~\bibnamefont {Cheng}}, \bibinfo {author} {\bibfnamefont {J.~W.}\ \bibnamefont {Haus}},\ and\ \bibinfo {author} {\bibfnamefont {Q.}~\bibnamefont {Zhan}},\ }\bibfield  {title} {\bibinfo {title} {Propagation of vector vortex beams through a turbulent atmosphere},\ }\href {https://doi.org/10.1364/OE.17.017829} {\bibfield  {journal} {\bibinfo  {journal} {Optics Express}\ }\textbf {\bibinfo {volume} {17}},\ \bibinfo {pages} {17829} (\bibinfo {year} {2009})}\BibitemShut {NoStop}%
\bibitem [{\citenamefont {Roxworthy}\ and\ \citenamefont {Toussaint}(2010)}]{roxworthy2010}%
  \BibitemOpen
  \bibfield  {author} {\bibinfo {author} {\bibfnamefont {B.~J.}\ \bibnamefont {Roxworthy}}\ and\ \bibinfo {author} {\bibfnamefont {K.~C.}\ \bibnamefont {Toussaint}},\ }\bibfield  {title} {\bibinfo {title} {Optical trapping with {$\pi$}-phase cylindrical vector beams},\ }\href {https://doi.org/10.1088/1367-2630/12/7/073012} {\bibfield  {journal} {\bibinfo  {journal} {New Journal of Physics}\ }\textbf {\bibinfo {volume} {12}},\ \bibinfo {pages} {073012} (\bibinfo {year} {2010})}\BibitemShut {NoStop}%
\bibitem [{\citenamefont {Fatemi}(2011)}]{fatemi2011}%
  \BibitemOpen
  \bibfield  {author} {\bibinfo {author} {\bibfnamefont {F.~K.}\ \bibnamefont {Fatemi}},\ }\bibfield  {title} {\bibinfo {title} {Cylindrical vector beams for rapid polarization-dependent measurements in atomic systems},\ }\href {https://doi.org/10.1364/OE.19.025143} {\bibfield  {journal} {\bibinfo  {journal} {Optics Express}\ }\textbf {\bibinfo {volume} {19}},\ \bibinfo {pages} {25143} (\bibinfo {year} {2011})}\BibitemShut {NoStop}%
\bibitem [{\citenamefont {Neugebauer}\ \emph {et~al.}(2014)\citenamefont {Neugebauer}, \citenamefont {Bauer}, \citenamefont {Banzer},\ and\ \citenamefont {Leuchs}}]{neugebauer2014}%
  \BibitemOpen
  \bibfield  {author} {\bibinfo {author} {\bibfnamefont {M.}~\bibnamefont {Neugebauer}}, \bibinfo {author} {\bibfnamefont {T.}~\bibnamefont {Bauer}}, \bibinfo {author} {\bibfnamefont {P.}~\bibnamefont {Banzer}},\ and\ \bibinfo {author} {\bibfnamefont {G.}~\bibnamefont {Leuchs}},\ }\bibfield  {title} {\bibinfo {title} {Polarization {{Tailored Light Driven Directional Optical Nanobeacon}}},\ }\href {https://doi.org/10.1021/nl5003526} {\bibfield  {journal} {\bibinfo  {journal} {Nano Letters}\ }\textbf {\bibinfo {volume} {14}},\ \bibinfo {pages} {2546} (\bibinfo {year} {2014})}\BibitemShut {NoStop}%
\bibitem [{\citenamefont {Parigi}\ \emph {et~al.}(2015)\citenamefont {Parigi}, \citenamefont {D'Ambrosio}, \citenamefont {Arnold}, \citenamefont {Marrucci}, \citenamefont {Sciarrino},\ and\ \citenamefont {Laurat}}]{parigi2015}%
  \BibitemOpen
  \bibfield  {author} {\bibinfo {author} {\bibfnamefont {V.}~\bibnamefont {Parigi}}, \bibinfo {author} {\bibfnamefont {V.}~\bibnamefont {D'Ambrosio}}, \bibinfo {author} {\bibfnamefont {C.}~\bibnamefont {Arnold}}, \bibinfo {author} {\bibfnamefont {L.}~\bibnamefont {Marrucci}}, \bibinfo {author} {\bibfnamefont {F.}~\bibnamefont {Sciarrino}},\ and\ \bibinfo {author} {\bibfnamefont {J.}~\bibnamefont {Laurat}},\ }\bibfield  {title} {\bibinfo {title} {Storage and retrieval of vector beams of light in a multiple-degree-of-freedom quantum memory},\ }\href {https://doi.org/10.1038/ncomms8706} {\bibfield  {journal} {\bibinfo  {journal} {Nature Communications}\ }\textbf {\bibinfo {volume} {6}},\ \bibinfo {pages} {7706} (\bibinfo {year} {2015})}\BibitemShut {NoStop}%
\bibitem [{\citenamefont {Yuan}\ \emph {et~al.}(2017)\citenamefont {Yuan}, \citenamefont {Lei}, \citenamefont {Gao}, \citenamefont {Weng}, \citenamefont {Du},\ and\ \citenamefont {Yuan}}]{yuan2017}%
  \BibitemOpen
  \bibfield  {author} {\bibinfo {author} {\bibfnamefont {Y.}~\bibnamefont {Yuan}}, \bibinfo {author} {\bibfnamefont {T.}~\bibnamefont {Lei}}, \bibinfo {author} {\bibfnamefont {S.}~\bibnamefont {Gao}}, \bibinfo {author} {\bibfnamefont {X.}~\bibnamefont {Weng}}, \bibinfo {author} {\bibfnamefont {L.}~\bibnamefont {Du}},\ and\ \bibinfo {author} {\bibfnamefont {X.}~\bibnamefont {Yuan}},\ }\bibfield  {title} {\bibinfo {title} {The {{Orbital Angular Momentum Spreading}} for {{Cylindrical Vector Beams}} in {{Turbulent Atmosphere}}},\ }\href {https://doi.org/10.1109/JPHOT.2017.2683499} {\bibfield  {journal} {\bibinfo  {journal} {IEEE Photonics Journal}\ }\textbf {\bibinfo {volume} {9}},\ \bibinfo {pages} {1} (\bibinfo {year} {2017})}\BibitemShut {NoStop}%
\bibitem [{\citenamefont {Yuan}\ \emph {et~al.}(2022)\citenamefont {Yuan}, \citenamefont {Xiao}, \citenamefont {Liu}, \citenamefont {Fu}, \citenamefont {Qu}, \citenamefont {Gbur},\ and\ \citenamefont {Cai}}]{yuan2022}%
  \BibitemOpen
  \bibfield  {author} {\bibinfo {author} {\bibfnamefont {Y.}~\bibnamefont {Yuan}}, \bibinfo {author} {\bibfnamefont {X.}~\bibnamefont {Xiao}}, \bibinfo {author} {\bibfnamefont {D.}~\bibnamefont {Liu}}, \bibinfo {author} {\bibfnamefont {P.}~\bibnamefont {Fu}}, \bibinfo {author} {\bibfnamefont {J.}~\bibnamefont {Qu}}, \bibinfo {author} {\bibfnamefont {G.}~\bibnamefont {Gbur}},\ and\ \bibinfo {author} {\bibfnamefont {Y.}~\bibnamefont {Cai}},\ }\bibfield  {title} {\bibinfo {title} {Mitigating orbital angular momentum crosstalk in an optical communication uplink channel using cylindrical vector beams},\ }\href {https://doi.org/10.1080/17455030.2022.2053609} {\bibfield  {journal} {\bibinfo  {journal} {Waves in Random and Complex Media}\ }\textbf {\bibinfo {volume} {0}},\ \bibinfo {pages} {1} (\bibinfo {year} {2022})}\BibitemShut {NoStop}%
\bibitem [{\citenamefont {Cheng}\ \emph {et~al.}(2023)\citenamefont {Cheng}, \citenamefont {Dong}, \citenamefont {Shi}, \citenamefont {Mohammed}, \citenamefont {Guo}, \citenamefont {Yi}, \citenamefont {Wang},\ and\ \citenamefont {Li}}]{cheng2023}%
  \BibitemOpen
  \bibfield  {author} {\bibinfo {author} {\bibfnamefont {M.}~\bibnamefont {Cheng}}, \bibinfo {author} {\bibfnamefont {K.}~\bibnamefont {Dong}}, \bibinfo {author} {\bibfnamefont {C.}~\bibnamefont {Shi}}, \bibinfo {author} {\bibfnamefont {A.-A. H.~T.}\ \bibnamefont {Mohammed}}, \bibinfo {author} {\bibfnamefont {L.}~\bibnamefont {Guo}}, \bibinfo {author} {\bibfnamefont {X.}~\bibnamefont {Yi}}, \bibinfo {author} {\bibfnamefont {P.}~\bibnamefont {Wang}},\ and\ \bibinfo {author} {\bibfnamefont {J.}~\bibnamefont {Li}},\ }\bibfield  {title} {\bibinfo {title} {Enhancing {{Performance}} of {{Air}}\textendash{{Ground OAM Communication System Utilizing Vector Vortex Beams}} in the {{Atmosphere}}},\ }\href {https://doi.org/10.3390/photonics10010041} {\bibfield  {journal} {\bibinfo  {journal} {Photonics}\ }\textbf {\bibinfo {volume} {10}},\ \bibinfo {pages} {41} (\bibinfo {year} {2023})}\BibitemShut {NoStop}%
\bibitem [{\citenamefont {Ndagano}\ \emph {et~al.}(2018)\citenamefont {Ndagano}, \citenamefont {Nape}, \citenamefont {Cox}, \citenamefont {{Rosales-Guzman}},\ and\ \citenamefont {Forbes}}]{ndagano2018}%
  \BibitemOpen
  \bibfield  {author} {\bibinfo {author} {\bibfnamefont {B.}~\bibnamefont {Ndagano}}, \bibinfo {author} {\bibfnamefont {I.}~\bibnamefont {Nape}}, \bibinfo {author} {\bibfnamefont {M.~A.}\ \bibnamefont {Cox}}, \bibinfo {author} {\bibfnamefont {C.}~\bibnamefont {{Rosales-Guzman}}},\ and\ \bibinfo {author} {\bibfnamefont {A.}~\bibnamefont {Forbes}},\ }\bibfield  {title} {\bibinfo {title} {Creation and {{Detection}} of {{Vector Vortex Modes}} for {{Classical}} and {{Quantum Communication}}},\ }\href {https://doi.org/10.1109/JLT.2017.2766760} {\bibfield  {journal} {\bibinfo  {journal} {Journal of Lightwave Technology}\ }\textbf {\bibinfo {volume} {36}},\ \bibinfo {pages} {292} (\bibinfo {year} {2018})}\BibitemShut {NoStop}%
\bibitem [{\citenamefont {Wang}(2016)}]{wang2016}%
  \BibitemOpen
  \bibfield  {author} {\bibinfo {author} {\bibfnamefont {J.}~\bibnamefont {Wang}},\ }\bibfield  {title} {\bibinfo {title} {Advances in communications using optical vortices},\ }\href {https://doi.org/10.1364/PRJ.4.000B14} {\bibfield  {journal} {\bibinfo  {journal} {Photonics Research}\ }\textbf {\bibinfo {volume} {4}},\ \bibinfo {pages} {B14} (\bibinfo {year} {2016})}\BibitemShut {NoStop}%
\bibitem [{\citenamefont {Wang}(2017)}]{wang2017}%
  \BibitemOpen
  \bibfield  {author} {\bibinfo {author} {\bibfnamefont {J.}~\bibnamefont {Wang}},\ }\bibfield  {title} {\bibinfo {title} {Data information transfer using complex optical fields: A review and perspective ({{Invited Paper}})},\ }\href@noop {} {\bibfield  {journal} {\bibinfo  {journal} {Chinese Optics Letters}\ }\textbf {\bibinfo {volume} {15}},\ \bibinfo {pages} {030005} (\bibinfo {year} {2017})}\BibitemShut {NoStop}%
\bibitem [{\citenamefont {Chille}\ \emph {et~al.}(2016)\citenamefont {Chille}, \citenamefont {{Berg-Johansen}}, \citenamefont {Semmler}, \citenamefont {Banzer}, \citenamefont {Aiello}, \citenamefont {Leuchs},\ and\ \citenamefont {Marquardt}}]{chille2016}%
  \BibitemOpen
  \bibfield  {author} {\bibinfo {author} {\bibfnamefont {V.}~\bibnamefont {Chille}}, \bibinfo {author} {\bibfnamefont {S.}~\bibnamefont {{Berg-Johansen}}}, \bibinfo {author} {\bibfnamefont {M.}~\bibnamefont {Semmler}}, \bibinfo {author} {\bibfnamefont {P.}~\bibnamefont {Banzer}}, \bibinfo {author} {\bibfnamefont {A.}~\bibnamefont {Aiello}}, \bibinfo {author} {\bibfnamefont {G.}~\bibnamefont {Leuchs}},\ and\ \bibinfo {author} {\bibfnamefont {C.}~\bibnamefont {Marquardt}},\ }\bibfield  {title} {\bibinfo {title} {Experimental generation of amplitude squeezed vector beams},\ }\href {https://doi.org/10.1364/OE.24.012385} {\bibfield  {journal} {\bibinfo  {journal} {Optics Express}\ }\textbf {\bibinfo {volume} {24}},\ \bibinfo {pages} {12385} (\bibinfo {year} {2016})}\BibitemShut {NoStop}%
\bibitem [{\citenamefont {Vallone}\ \emph {et~al.}(2014)\citenamefont {Vallone}, \citenamefont {D'Ambrosio}, \citenamefont {Sponselli}, \citenamefont {Slussarenko}, \citenamefont {Marrucci}, \citenamefont {Sciarrino},\ and\ \citenamefont {Villoresi}}]{vallone2014}%
  \BibitemOpen
  \bibfield  {author} {\bibinfo {author} {\bibfnamefont {G.}~\bibnamefont {Vallone}}, \bibinfo {author} {\bibfnamefont {V.}~\bibnamefont {D'Ambrosio}}, \bibinfo {author} {\bibfnamefont {A.}~\bibnamefont {Sponselli}}, \bibinfo {author} {\bibfnamefont {S.}~\bibnamefont {Slussarenko}}, \bibinfo {author} {\bibfnamefont {L.}~\bibnamefont {Marrucci}}, \bibinfo {author} {\bibfnamefont {F.}~\bibnamefont {Sciarrino}},\ and\ \bibinfo {author} {\bibfnamefont {P.}~\bibnamefont {Villoresi}},\ }\bibfield  {title} {\bibinfo {title} {Free-{{Space Quantum Key Distribution}} by {{Rotation-Invariant Twisted Photons}}},\ }\href {https://doi.org/10.1103/PhysRevLett.113.060503} {\bibfield  {journal} {\bibinfo  {journal} {Physical Review Letters}\ }\textbf {\bibinfo {volume} {113}},\ \bibinfo {pages} {060503} (\bibinfo {year} {2014})}\BibitemShut {NoStop}%
\bibitem [{\citenamefont {D'Ambrosio}\ \emph {et~al.}(2012)\citenamefont {D'Ambrosio}, \citenamefont {Nagali}, \citenamefont {Walborn}, \citenamefont {Aolita}, \citenamefont {Slussarenko}, \citenamefont {Marrucci},\ and\ \citenamefont {Sciarrino}}]{dambrosio2012a}%
  \BibitemOpen
  \bibfield  {author} {\bibinfo {author} {\bibfnamefont {V.}~\bibnamefont {D'Ambrosio}}, \bibinfo {author} {\bibfnamefont {E.}~\bibnamefont {Nagali}}, \bibinfo {author} {\bibfnamefont {S.~P.}\ \bibnamefont {Walborn}}, \bibinfo {author} {\bibfnamefont {L.}~\bibnamefont {Aolita}}, \bibinfo {author} {\bibfnamefont {S.}~\bibnamefont {Slussarenko}}, \bibinfo {author} {\bibfnamefont {L.}~\bibnamefont {Marrucci}},\ and\ \bibinfo {author} {\bibfnamefont {F.}~\bibnamefont {Sciarrino}},\ }\bibfield  {title} {\bibinfo {title} {Complete experimental toolbox for alignment-free quantum communication},\ }\href {https://doi.org/10.1038/ncomms1951} {\bibfield  {journal} {\bibinfo  {journal} {Nature Communications}\ }\textbf {\bibinfo {volume} {3}},\ \bibinfo {pages} {961} (\bibinfo {year} {2012})}\BibitemShut {NoStop}%
\bibitem [{\citenamefont {D'Ambrosio}\ \emph {et~al.}(2016)\citenamefont {D'Ambrosio}, \citenamefont {Carvacho}, \citenamefont {Graffitti}, \citenamefont {Vitelli}, \citenamefont {Piccirillo}, \citenamefont {Marrucci},\ and\ \citenamefont {Sciarrino}}]{dambrosio2016}%
  \BibitemOpen
  \bibfield  {author} {\bibinfo {author} {\bibfnamefont {V.}~\bibnamefont {D'Ambrosio}}, \bibinfo {author} {\bibfnamefont {G.}~\bibnamefont {Carvacho}}, \bibinfo {author} {\bibfnamefont {F.}~\bibnamefont {Graffitti}}, \bibinfo {author} {\bibfnamefont {C.}~\bibnamefont {Vitelli}}, \bibinfo {author} {\bibfnamefont {B.}~\bibnamefont {Piccirillo}}, \bibinfo {author} {\bibfnamefont {L.}~\bibnamefont {Marrucci}},\ and\ \bibinfo {author} {\bibfnamefont {F.}~\bibnamefont {Sciarrino}},\ }\bibfield  {title} {\bibinfo {title} {Entangled vector vortex beams},\ }\href {https://doi.org/10.1103/PhysRevA.94.030304} {\bibfield  {journal} {\bibinfo  {journal} {Physical Review A}\ }\textbf {\bibinfo {volume} {94}},\ \bibinfo {pages} {030304} (\bibinfo {year} {2016})}\BibitemShut {NoStop}%
\bibitem [{\citenamefont {Berkhout}\ \emph {et~al.}(2010)\citenamefont {Berkhout}, \citenamefont {Lavery}, \citenamefont {Courtial}, \citenamefont {Beijersbergen},\ and\ \citenamefont {Padgett}}]{Berkhout2010}%
  \BibitemOpen
  \bibfield  {author} {\bibinfo {author} {\bibfnamefont {G.~C.}\ \bibnamefont {Berkhout}}, \bibinfo {author} {\bibfnamefont {M.~P.}\ \bibnamefont {Lavery}}, \bibinfo {author} {\bibfnamefont {J.}~\bibnamefont {Courtial}}, \bibinfo {author} {\bibfnamefont {M.~W.}\ \bibnamefont {Beijersbergen}},\ and\ \bibinfo {author} {\bibfnamefont {M.~J.}\ \bibnamefont {Padgett}},\ }\bibfield  {title} {\bibinfo {title} {Efficient sorting of orbital angular momentum states of light},\ }\href@noop {} {\bibfield  {journal} {\bibinfo  {journal} {Physical review letters}\ }\textbf {\bibinfo {volume} {105}},\ \bibinfo {pages} {153601} (\bibinfo {year} {2010})}\BibitemShut {NoStop}%
\bibitem [{\citenamefont {Wei}\ \emph {et~al.}(2020)\citenamefont {Wei}, \citenamefont {Earl}, \citenamefont {Lin}, \citenamefont {Kou},\ and\ \citenamefont {Yuan}}]{wei2020}%
  \BibitemOpen
  \bibfield  {author} {\bibinfo {author} {\bibfnamefont {S.}~\bibnamefont {Wei}}, \bibinfo {author} {\bibfnamefont {S.~K.}\ \bibnamefont {Earl}}, \bibinfo {author} {\bibfnamefont {J.}~\bibnamefont {Lin}}, \bibinfo {author} {\bibfnamefont {S.~S.}\ \bibnamefont {Kou}},\ and\ \bibinfo {author} {\bibfnamefont {X.-C.}\ \bibnamefont {Yuan}},\ }\bibfield  {title} {\bibinfo {title} {Active sorting of orbital angular momentum states of light with a cascaded tunable resonator},\ }\href {https://doi.org/10.1038/s41377-020-0243-x} {\bibfield  {journal} {\bibinfo  {journal} {Light: Science \& Applications}\ }\textbf {\bibinfo {volume} {9}},\ \bibinfo {pages} {10} (\bibinfo {year} {2020})}\BibitemShut {NoStop}%
\bibitem [{\citenamefont {dos Santos}\ \emph {et~al.}(2021)\citenamefont {dos Santos}, \citenamefont {Salles}, \citenamefont {Damaceno}, \citenamefont {Menezes}, \citenamefont {Corso}, \citenamefont {Martinelli}, \citenamefont {Ribeiro},\ and\ \citenamefont {de~Ara{\'u}jo}}]{Santos2021}%
  \BibitemOpen
  \bibfield  {author} {\bibinfo {author} {\bibfnamefont {G.}~\bibnamefont {dos Santos}}, \bibinfo {author} {\bibfnamefont {D.~C.~d.}\ \bibnamefont {Salles}}, \bibinfo {author} {\bibfnamefont {M.~G.}\ \bibnamefont {Damaceno}}, \bibinfo {author} {\bibfnamefont {B.~T.~d.}\ \bibnamefont {Menezes}}, \bibinfo {author} {\bibfnamefont {C.}~\bibnamefont {Corso}}, \bibinfo {author} {\bibfnamefont {M.}~\bibnamefont {Martinelli}}, \bibinfo {author} {\bibfnamefont {P.~S.}\ \bibnamefont {Ribeiro}},\ and\ \bibinfo {author} {\bibfnamefont {R.~M.}\ \bibnamefont {de~Ara{\'u}jo}},\ }\bibfield  {title} {\bibinfo {title} {Decomposing spatial mode superpositions with a triangular optical cavity},\ }\href@noop {} {\bibfield  {journal} {\bibinfo  {journal} {Physical Review Applied}\ }\textbf {\bibinfo {volume} {16}},\ \bibinfo {pages} {034008} (\bibinfo {year} {2021})}\BibitemShut {NoStop}%
\bibitem [{\citenamefont {Gouy}(1890)}]{Gouy1890}%
  \BibitemOpen
  \bibfield  {author} {\bibinfo {author} {\bibfnamefont {L.~G.}\ \bibnamefont {Gouy}},\ }\href@noop {} {\emph {\bibinfo {title} {Sur une propri{\'e}t{\'e} nouvelle des ondes lumineuses}}}\ (\bibinfo  {publisher} {Gauthier-Villars},\ \bibinfo {address} {Paris},\ \bibinfo {year} {1890})\BibitemShut {NoStop}%
\bibitem [{\citenamefont {Shibiao}\ \emph {et~al.}(2020)\citenamefont {Shibiao}, \citenamefont {Earl}, \citenamefont {Jiao}, \citenamefont {Kou},\ and\ \citenamefont {Xiao-Cong}}]{Shibiao2020}%
  \BibitemOpen
  \bibfield  {author} {\bibinfo {author} {\bibfnamefont {W.}~\bibnamefont {Shibiao}}, \bibinfo {author} {\bibfnamefont {S.~K.}\ \bibnamefont {Earl}}, \bibinfo {author} {\bibfnamefont {L.}~\bibnamefont {Jiao}}, \bibinfo {author} {\bibfnamefont {S.~S.}\ \bibnamefont {Kou}},\ and\ \bibinfo {author} {\bibfnamefont {Y.}~\bibnamefont {Xiao-Cong}},\ }\bibfield  {title} {\bibinfo {title} {Active sorting of orbital angular momentum states of light with a cascaded tunable resonator},\ }\href@noop {} {\bibfield  {journal} {\bibinfo  {journal} {Light: Science and Applications}\ }\textbf {\bibinfo {volume} {9}} (\bibinfo {year} {2020})}\BibitemShut {NoStop}%
\bibitem [{\citenamefont {Sasada}\ and\ \citenamefont {Okamoto}(2003)}]{Sasada2003}%
  \BibitemOpen
  \bibfield  {author} {\bibinfo {author} {\bibfnamefont {H.}~\bibnamefont {Sasada}}\ and\ \bibinfo {author} {\bibfnamefont {M.}~\bibnamefont {Okamoto}},\ }\bibfield  {title} {\bibinfo {title} {Transverse-mode beam splitter of a light beam and its application to quantum cryptography},\ }\href@noop {} {\bibfield  {journal} {\bibinfo  {journal} {Physical Review A}\ }\textbf {\bibinfo {volume} {68}},\ \bibinfo {pages} {012323} (\bibinfo {year} {2003})}\BibitemShut {NoStop}%
\bibitem [{\citenamefont {Zhan}(2009)}]{zhan2009}%
  \BibitemOpen
  \bibfield  {author} {\bibinfo {author} {\bibfnamefont {Q.}~\bibnamefont {Zhan}},\ }\bibfield  {title} {\bibinfo {title} {Cylindrical vector beams: from mathematical concepts to applications},\ }\href {https://doi.org/10.1364/AOP.1.000001} {\bibfield  {journal} {\bibinfo  {journal} {Advances in Optics and Photonics}\ }\textbf {\bibinfo {volume} {1}},\ \bibinfo {pages} {1} (\bibinfo {year} {2009})}\BibitemShut {NoStop}%
\bibitem [{\citenamefont {Pancharatnam}(1956)}]{Pancharatnam1956}%
  \BibitemOpen
  \bibfield  {author} {\bibinfo {author} {\bibfnamefont {S.}~\bibnamefont {Pancharatnam}},\ }\bibfield  {title} {\bibinfo {title} {Generalized theory of interference and its applications},\ }\href {https://doi.org/10.1007/BF03046050} {\bibfield  {journal} {\bibinfo  {journal} {Proc. Indian Acad. Sci.}\ }\textbf {\bibinfo {volume} {44}},\ \bibinfo {pages} {247} (\bibinfo {year} {1956})}\BibitemShut {NoStop}%
\bibitem [{\citenamefont {Berry}(1984)}]{Berry1984}%
  \BibitemOpen
  \bibfield  {author} {\bibinfo {author} {\bibfnamefont {M.~V.}\ \bibnamefont {Berry}},\ }\bibfield  {title} {\bibinfo {title} {Quantal phase factors accompanying adiabatic changes},\ }\href {http://www.jstor.org/stable/2397741} {\bibfield  {journal} {\bibinfo  {journal} {Proceedings of the Royal Society of London. Series A, Mathematical and Physical Sciences}\ }\textbf {\bibinfo {volume} {392}},\ \bibinfo {pages} {45} (\bibinfo {year} {1984})}\BibitemShut {NoStop}%
\bibitem [{\citenamefont {Mukunda}\ and\ \citenamefont {Simon}(1993)}]{Mukunda1993}%
  \BibitemOpen
  \bibfield  {author} {\bibinfo {author} {\bibfnamefont {N.}~\bibnamefont {Mukunda}}\ and\ \bibinfo {author} {\bibfnamefont {R.}~\bibnamefont {Simon}},\ }\bibfield  {title} {\bibinfo {title} {Quantum kinematic approach to the geometric phase. i. general formalism},\ }\href {https://doi.org/https://doi.org/10.1006/aphy.1993.1093} {\bibfield  {journal} {\bibinfo  {journal} {Annals of Physics}\ }\textbf {\bibinfo {volume} {228}},\ \bibinfo {pages} {205} (\bibinfo {year} {1993})}\BibitemShut {NoStop}%
\bibitem [{\citenamefont {Jones}\ \emph {et~al.}(2000)\citenamefont {Jones}, \citenamefont {Vedral}, \citenamefont {Ekert},\ and\ \citenamefont {Castagnoli}}]{Jones2000}%
  \BibitemOpen
  \bibfield  {author} {\bibinfo {author} {\bibfnamefont {J.~A.}\ \bibnamefont {Jones}}, \bibinfo {author} {\bibfnamefont {V.}~\bibnamefont {Vedral}}, \bibinfo {author} {\bibfnamefont {A.}~\bibnamefont {Ekert}},\ and\ \bibinfo {author} {\bibfnamefont {G.}~\bibnamefont {Castagnoli}},\ }\bibfield  {title} {\bibinfo {title} {Geometric quantum computation using nuclear magnetic resonance},\ }\href {https://doi.org/10.1038/35002528} {\bibfield  {journal} {\bibinfo  {journal} {Nature}\ }\textbf {\bibinfo {volume} {403}},\ \bibinfo {pages} {869} (\bibinfo {year} {2000})}\BibitemShut {NoStop}%
\bibitem [{\citenamefont {Duan}\ \emph {et~al.}(2001)\citenamefont {Duan}, \citenamefont {Cirac},\ and\ \citenamefont {Zoller}}]{Duan2001}%
  \BibitemOpen
  \bibfield  {author} {\bibinfo {author} {\bibfnamefont {L.~M.}\ \bibnamefont {Duan}}, \bibinfo {author} {\bibfnamefont {J.~I.}\ \bibnamefont {Cirac}},\ and\ \bibinfo {author} {\bibfnamefont {P.}~\bibnamefont {Zoller}},\ }\bibfield  {title} {\bibinfo {title} {Geometric manipulation of trapped ions for quantum computation},\ }\href {https://doi.org/10.1126/science.1058835} {\bibfield  {journal} {\bibinfo  {journal} {Science}\ }\textbf {\bibinfo {volume} {292}},\ \bibinfo {pages} {1965} (\bibinfo {year} {2001})}\BibitemShut {NoStop}%
\bibitem [{\citenamefont {{van Enk}}(1993)}]{VanEnk1993}%
  \BibitemOpen
  \bibfield  {author} {\bibinfo {author} {\bibfnamefont {S.}~\bibnamefont {{van Enk}}},\ }\bibfield  {title} {\bibinfo {title} {Geometric phase, transformations of gaussian light beams and angular momentum transfer},\ }\href {https://doi.org/https://doi.org/10.1016/0030-4018(93)90472-H} {\bibfield  {journal} {\bibinfo  {journal} {Optics Communications}\ }\textbf {\bibinfo {volume} {102}},\ \bibinfo {pages} {59} (\bibinfo {year} {1993})}\BibitemShut {NoStop}%
\bibitem [{\citenamefont {Galvez}\ \emph {et~al.}(2003)\citenamefont {Galvez}, \citenamefont {Crawford}, \citenamefont {Sztul}, \citenamefont {Pysher}, \citenamefont {Haglin},\ and\ \citenamefont {Williams}}]{Galvez2003}%
  \BibitemOpen
  \bibfield  {author} {\bibinfo {author} {\bibfnamefont {E.~J.}\ \bibnamefont {Galvez}}, \bibinfo {author} {\bibfnamefont {P.~R.}\ \bibnamefont {Crawford}}, \bibinfo {author} {\bibfnamefont {H.~I.}\ \bibnamefont {Sztul}}, \bibinfo {author} {\bibfnamefont {M.~J.}\ \bibnamefont {Pysher}}, \bibinfo {author} {\bibfnamefont {P.~J.}\ \bibnamefont {Haglin}},\ and\ \bibinfo {author} {\bibfnamefont {R.~E.}\ \bibnamefont {Williams}},\ }\bibfield  {title} {\bibinfo {title} {Geometric phase associated with mode transformations of optical beams bearing orbital angular momentum},\ }\href {https://doi.org/10.1103/PhysRevLett.90.203901} {\bibfield  {journal} {\bibinfo  {journal} {Phys. Rev. Lett.}\ }\textbf {\bibinfo {volume} {90}},\ \bibinfo {pages} {203901} (\bibinfo {year} {2003})}\BibitemShut {NoStop}%
\bibitem [{\citenamefont {D\'ecamps}\ \emph {et~al.}(2017)\citenamefont {D\'ecamps}, \citenamefont {Gauguet}, \citenamefont {Vigu\'e},\ and\ \citenamefont {B\"uchner}}]{Decamps2017}%
  \BibitemOpen
  \bibfield  {author} {\bibinfo {author} {\bibfnamefont {B.}~\bibnamefont {D\'ecamps}}, \bibinfo {author} {\bibfnamefont {A.}~\bibnamefont {Gauguet}}, \bibinfo {author} {\bibfnamefont {J.}~\bibnamefont {Vigu\'e}},\ and\ \bibinfo {author} {\bibfnamefont {M.}~\bibnamefont {B\"uchner}},\ }\bibfield  {title} {\bibinfo {title} {Pancharatnam phase: A tool for atom optics},\ }\href {https://doi.org/10.1103/PhysRevA.96.013624} {\bibfield  {journal} {\bibinfo  {journal} {Phys. Rev. A}\ }\textbf {\bibinfo {volume} {96}},\ \bibinfo {pages} {013624} (\bibinfo {year} {2017})}\BibitemShut {NoStop}%
\bibitem [{\citenamefont {Djevahirdjian}\ \emph {et~al.}(2020)\citenamefont {Djevahirdjian}, \citenamefont {M{\'e}jean},\ and\ \citenamefont {Romanini}}]{djevahirdjian2020}%
  \BibitemOpen
  \bibfield  {author} {\bibinfo {author} {\bibfnamefont {L.}~\bibnamefont {Djevahirdjian}}, \bibinfo {author} {\bibfnamefont {G.}~\bibnamefont {M{\'e}jean}},\ and\ \bibinfo {author} {\bibfnamefont {D.}~\bibnamefont {Romanini}},\ }\bibfield  {title} {\bibinfo {title} {Gouy phase shift measurement in a high-finesse cavity by optical feedback frequency locking},\ }\href {https://doi.org/10.1088/1361-6501/ab501b} {\bibfield  {journal} {\bibinfo  {journal} {Measurement Science and Technology}\ }\textbf {\bibinfo {volume} {31}},\ \bibinfo {pages} {035013} (\bibinfo {year} {2020})}\BibitemShut {NoStop}%
\bibitem [{\citenamefont {Aolita}\ and\ \citenamefont {Walborn}(2007)}]{Aolita2007}%
  \BibitemOpen
  \bibfield  {author} {\bibinfo {author} {\bibfnamefont {L.}~\bibnamefont {Aolita}}\ and\ \bibinfo {author} {\bibfnamefont {S.}~\bibnamefont {Walborn}},\ }\bibfield  {title} {\bibinfo {title} {Quantum communication without alignment using multiple-qubit single-photon states},\ }\href@noop {} {\bibfield  {journal} {\bibinfo  {journal} {Physical review letters}\ }\textbf {\bibinfo {volume} {98}},\ \bibinfo {pages} {100501} (\bibinfo {year} {2007})}\BibitemShut {NoStop}%
\bibitem [{\citenamefont {Souza}\ \emph {et~al.}(2008)\citenamefont {Souza}, \citenamefont {Borges}, \citenamefont {Khoury}, \citenamefont {Huguenin}, \citenamefont {Aolita},\ and\ \citenamefont {Walborn}}]{Souza2008}%
  \BibitemOpen
  \bibfield  {author} {\bibinfo {author} {\bibfnamefont {C.}~\bibnamefont {Souza}}, \bibinfo {author} {\bibfnamefont {C.}~\bibnamefont {Borges}}, \bibinfo {author} {\bibfnamefont {A.}~\bibnamefont {Khoury}}, \bibinfo {author} {\bibfnamefont {J.}~\bibnamefont {Huguenin}}, \bibinfo {author} {\bibfnamefont {L.}~\bibnamefont {Aolita}},\ and\ \bibinfo {author} {\bibfnamefont {S.}~\bibnamefont {Walborn}},\ }\bibfield  {title} {\bibinfo {title} {Quantum key distribution without a shared reference frame},\ }\href@noop {} {\bibfield  {journal} {\bibinfo  {journal} {Physical Review A}\ }\textbf {\bibinfo {volume} {77}},\ \bibinfo {pages} {032345} (\bibinfo {year} {2008})}\BibitemShut {NoStop}%
\end{thebibliography}%

\section*{Acknowledgements (not compulsory)}

We thank Prof. Marcelo Martinelli for providing us with the laser source. We also acknowledge funding from the Brazilian funding agencies, Conselho Nacional de Desenvolvimento Tecnol\'ogico (CNPq), Funda\c c\~{a}o de Amparo \`{a} Pesquisa e Inova\c{c}\~{a}o do Estado de Santa Catarina (FAPESC), Coordena\c c\~{a}o de Aperfei\c coamento de Pessoal de N\'ivel Superior - Brasil (CAPES) and Instituto Nacional de Ci\^encia e Tecnologia de Informa\c c\~ao Qu\^antica (INCT/IQ 465469/2014-0). A.Z.K. acknowledges financial support from Fundação Carlos Chagas Filho de Amparo à Pesquisa do Estado do Rio de Janeiro (FAPERJ), and Fundação de Amparo à Pesquisa do
Estado de São Paulo (FAPESP, Grant No. 2021/06823-5).

\section*{Author contributions statement}

RMA, AZK and DCS conceived the experiment; LMR, LMF and GHS conducted the experiment; and analysed the results; PHSR provided financial support; JMK designed and printed essential pieces of the experiment. All authors contibuted writing and reviewing the manuscript. 

\section*{Additional information}

The authors declare no competing interests.

\end{document}